\documentclass[aps,prl,reprint]{revtex4-1}
\usepackage{blindtext}

\usepackage{mathrsfs}
\usepackage{amssymb,amsmath,mathtools}
\usepackage{amssymb}
\usepackage{dsfont}
\usepackage{graphicx}
\usepackage[usenames,dvipsnames,svgnames,x11names,table]{xcolor}
\usepackage{bm}
\usepackage{subfigure}
\usepackage{hyperref}
\usepackage[normalem]{ulem}
\usepackage{dcolumn}

\usepackage[utf8]{inputenc} 
\usepackage{xfrac}

\graphicspath{%
    {converted_graphics/}
    {Figuras/}
}
\begin{document}

\title{Interference effects due to nuclear motion of the hydrogen molecule}

\author{L. O. Santos}
\affiliation{CEFET/RJ, UnED Maria da Graça, Rio de Janeiro, RJ 25620-003, Brazil}
\author{Amanda Alencar}
\author{ I. Prazeres}
\author{ F. Impens}
\author{ C.~R.~de~Carvalho}
\author{N.~V.~de~Castro~Faria}
\author{Ginette Jalbert}
\email{ginette@if.ufrj.br\\}
\affiliation{Instituto de F\'\i sica, Universidade Federal do Rio de Janeiro, Cx. Postal 68528, Rio de Janeiro, RJ 21941-972, Brazil}

\begin{abstract}
We show that two-particle interferences can be used to probe the nuclear motion in a doubly-excited hydrogen molecule.  The dissociation of molecular hydrogen by electron impact involves several decay channels,  associated to different molecular rotational states, which produce quantum interferences in the detection of the atomic fragments. Thanks to the correlations between the angular momentum and vibrational states of the molecule, the fragments arising from each dissociation channel carry out a phase-shift which is a signature of the molecule rotation. These phase-shifts, which cannot be observed in a single-atom detection scheme, may be witnessed in realistic experimental conditions in a time-of-flight coincidence measurement. We analyse the interferences arising from the two lowest-energy rotational states of a para-hydrogen molecule. Our result shows the relevance of two-fragments correlations to track the molecular rotation.
\end{abstract}
 

\maketitle

Two-photon correlations are essential in the modern definition of coherence~\cite{Glauber}. In fact, two-photon interference experiments, such as the Hong-Ou-Mandel interference~\cite{HOMpaper}, have played a key role in the developments of quantum optics. Recently, two-particle interference has been performed involving atoms instead of photons in a Hong-Ou-Mandel type experiment~\cite{HOM-atoms}. Besides, two-photon correlations can provide  informations about the light source beyond those obtained with a single intensity measurement~\cite{RefKaiser}.  On the other hand, in the dissociation of a diatomic molecule, the source of the final two-particle state is the molecule itself. Connected to this, quantum interferences have been predicted, when there are at least two distinct dissociative states excited at the same energy~\cite{Beswick} and, for instance, have been measured for photodissociation of H$_2$ \cite{Glass-Maujean} and more recently for D$_2$ \cite{Mo}. 
In this paper it is shown that two-atom correlations may provide an insight into the nuclear motion that is unaccessible to single-atom measurements; more precisely, it is theoretically foreseen a interference pattern in the superposition of the wavefunctions of two H(2s) fragments, which emerge from the dissociation of H$_2$, related to different molecular rotational states. It must be emphasized that such two-atom interferences may be observed in a standard coincidence time-of-flight detection experiment similar to the setup reported in Ref.~\cite{J.Robert} for the production of metastable H(2s) hydrogen fragments.
The knowledge of the initial molecular state is important to design an experiment such as the one suggested in Ref.~\cite{carvalho15} which constitutes a potential
twin-atom source.

Long-lived metastable fragments can be obtained through the H(2s)+H(2s) dissociation channel of the doubly excited states of molecular hydrogen. The production of these doubly excited states by electron impact has been recently analysed~\cite{L.Santos2}. Multiple dissociative states correspond to different angular momentum contributions of the internuclear axis rotation. Hence, the distinct dissociation paths which produce quantum interferences are labelled by different rotational states of the molecule.

The effective molecular potential felt by the nuclei couples the angular momentum of the them to the vibrational state of the molecule. The phase-shifts imprinted on the atomic fragments by this effective molecular potential depend on the molecule rotation. These ``rotational'' phases cannot be observed in an experiment involving a single atomic detector. Their difference may nevertheless be measured in a coincidence time-of-flight detection experiment involving simultaneously the two H(2s) fragments produced in the dissociation.

In the Born-Oppenheimer approximation~(BOA),  following the Refs.~\cite{Bransden,assin}, the molecular wavefunction in the laboratory frame without spins is given by 
\begin{eqnarray}
\Psi_{mol}=\frac{1}{r}\sum_{J,M_J}c_{J,M_J}\psi_{el,\Lambda}\chi_{\nu,J}{Y}_J^{M_J}(\theta,\phi),
\label{eq1-corr}
\end{eqnarray}
where: {\bf J} = {\bf L}+{\bf N} is the total angular momentum; {\bf L} and {\bf N} stand for the electrons and nuclei angular momenta, respectively; $\Lambda$ and $M_J$ are the projections of {\bf L} and {\bf J} along the internuclear axis, respectively; $\psi_{el,\Lambda}$ is the electronic wavefunction; $\chi_{\nu,J}$ and $Y^J_{M_J}$ are the nuclei vibrational and rotational ones; and finally ($\theta,\phi$) are the angles of the nuclear axis in respect to the laboratory frame~\cite{Bransden,assin}.

The molecular wavefunction must fulfill symmetry requirements in compliance with the fermionic nature of its components.  In its ground state H$_2$ lies in the $^1\Sigma_g^+$ electronic state and it has been shown that the doubly excited states, which produces H(2s)+H(2s) fragments and cannot be reached by photon excitation, has the same electronic symmetry~\cite{L.Santos1}. Therefore, it leads to {\bf L }= 0, $\Lambda=0$ and {\bf J} ={\bf N}. It is worth mentioning that the electronic part is symmetric by the nuclei exchange. Considering $\widetilde{\Psi}_{mol}=\Psi_{mol}~\xi_{el}$ as the molecular wavefunction plus the electronic spin ($\xi_{el}$) and taking into account the nuclear spin, $\xi_N$, one has two distinct possibilities:
\begin{itemize}
\item $\Psi_{tot}^A = \widetilde{\Psi}_{mol}^S ~\xi_N^A$ (para-hydrogen); 
\item $\Psi_{tot}^A = \widetilde{\Psi}_{mol}^A\xi_N^S$ (ortho-hydrogen). 
\end{itemize}
Therefore, for the electronic state $^1\Sigma_g^+$, one has the wavefunction $\widetilde{\Psi}_{mol}^S$ for {\bf J} = {\bf N} even, and $\widetilde{\Psi}_{mol}^A$ for {\bf J} = {\bf N} odd.  We consider only the para-hydrogen configuration. Although in ordinary conditions a sample of H$_2$, in its ground state, has its ortho form three times as abundant as the para one, the choice of only considering the para-hydrogen corresponds nevertheless to a realistic experimental configuration as pure samples of para-hydrogen can be produced with well known techniques \cite{Leighton59, Wall74}. 

As the molecule nuclear spin state is unchanged by the electron collision, one can say that the global excited state here considered has the same symmetry as the initial ground state.

At room temperature, the ground state of the molecular hydrogen in thermal equilibrium, has few rotational levels significantly populated (up to N = 3). Moreover, rotational transitions are not likely to happen in electron impact collisions (especially at high electron impact energy range) \cite{OKSYUK}, then one can consider that only rotational transitions with small $\Delta N$ take place after collision. Thus, reminding that we are dealing with para-hydrogen, we shall assume that the initial rotational state is only $N=0$, leading to an excited state which is a linear combination of $N=0$ and $N=2$. Without the center-of-mass motion, one can write Eq.(\ref{eq1-corr}) as
\begin{eqnarray}
\psi_{mol} = \sum_{N \in \{0,2\} ,M_N} c_{N,M_N}\psi_{el,0}   
  \times \frac {\chi_{\nu,N}} {r} Y^N_{M_N}(\theta,\phi) \, .
\label{eq2-corr}
\end{eqnarray}
Instead of deriving explicitly the coefficients $c_{N,M_N}$ associated to the  collisional process, we shall focus on the asymptotic form of the wavefunction and look for the associated nuclei momentum distributions. 

As we are interesting in the outgoing fragments, only the outgoing contribution of the  asymptotic form of the vibrational wavefunction $\chi_{\nu,N}(r \rightarrow + \infty ) = e^{i (k r+2\delta_N-\frac{N\pi}{2})}$ is retained. The associated frequency reads $h\nu=\frac{\hbar^2 k^2}{2\mu}$, where $\mu$ is the reduced mass, $\hbar k$ is the relative momentum of the nuclei and $r$ its relative position. One of the signatures of the repulsive potential is contained in the phase-shift $\delta_N$, which can be obtained by treating the dissociation process as a time-reversed collisional process. 

A complete description in the laboratory frame requires to take into account the center-of-mass motion. In this sense, it is useful to project the nuclei momenta along the axes formed by the two detectors and to introduce the nuclei wave-vectors:
\begin{equation}
 k_{A}^{\pm} =\frac{K}{2}\pm k,
\hspace{.5cm} 
 k_{B}^{\pm} =\frac{K}{2}\mp k,
\label{kAkB}
\end{equation}
where $\hbar K $ corresponds to the center-of-mass and $\hbar k$ to the relative momentum of the two nuclei labeled by the letters $A,B$. The signals $+$ and $-$ corresponds to the direction of emission of the nuclei $A$ in the center-of-mass frame.  
Moreover, in order to obtain a proper description of the asymptotic wavefunction one has to take into account all the possible relative momenta allowed in the Franck-Condon region~\cite{Bransden}. The detection system can be designed to be sensitive exclusively to the fragments in the $|2s\rangle$ state~\cite{J.Robert}. Thus, in the asymptotic limit $r\rightarrow\infty$ , the molecular wavefunction corresponding to the dissociation of H$_2$ in the state $Q_2$ $^1\Sigma_g^+$
 takes the form
\begin{eqnarray}
\hspace{-.3cm} \Psi_{asy}  &&= \sum_{N,M_N}\int dKG(K)\int dkf_N(k)e^{-\frac{iN\pi}{2}}e^{i2\delta_N} \times \nonumber \\ \nonumber \\
& &\times \Bigl[ \Theta(r_B-r_A)e^{i(k_A^- r_A+k_B^-r_B)} \nonumber \\ \nonumber \\
&& +~ \Theta(r_A-r_B)e^{i(k_A^+ r_A+k_B^+r_B)} \Bigr] \times \nonumber \\ \nonumber \\
& &\times \Bigl(\vert 2s\rangle_{A}^{1}\vert 2s\rangle_{B}^{2} + \vert 2s\rangle_{A}^{2}\vert 2s\rangle_{B}^{1} \Bigr) Y_N^{M_N}(\theta,\phi)
\label{Psi-asymp}
\end{eqnarray}
where $G(K)$ is related to the CM moment distribution and $f_N(k)$ to the relative one, which is obtained from the reflection method~\cite{aline11}. Note that the function $\Theta(r_i-r_j)$ ($i$ and $j$ standing for $A$ and $B$) guarantees the correct signs of the exponentials arguments depending on which particle goes to left and to right. From now on we replace $Y_N^{M_N}(\theta,\phi)$ by $|N,M_N\rangle$. As we are dealing with thermal molecules whose velocities are null on average, we may consider null the molecule's center-of-mass moment, which corresponds to $G(K)=\delta(0)$; consequently, the moment of the fragments are opposite and of the same magnitude in the laboratory frame of reference as can be seen in Eq.~(\ref{kAkB}).

\begin{figure}[h]
\centering
\includegraphics[scale=0.2]{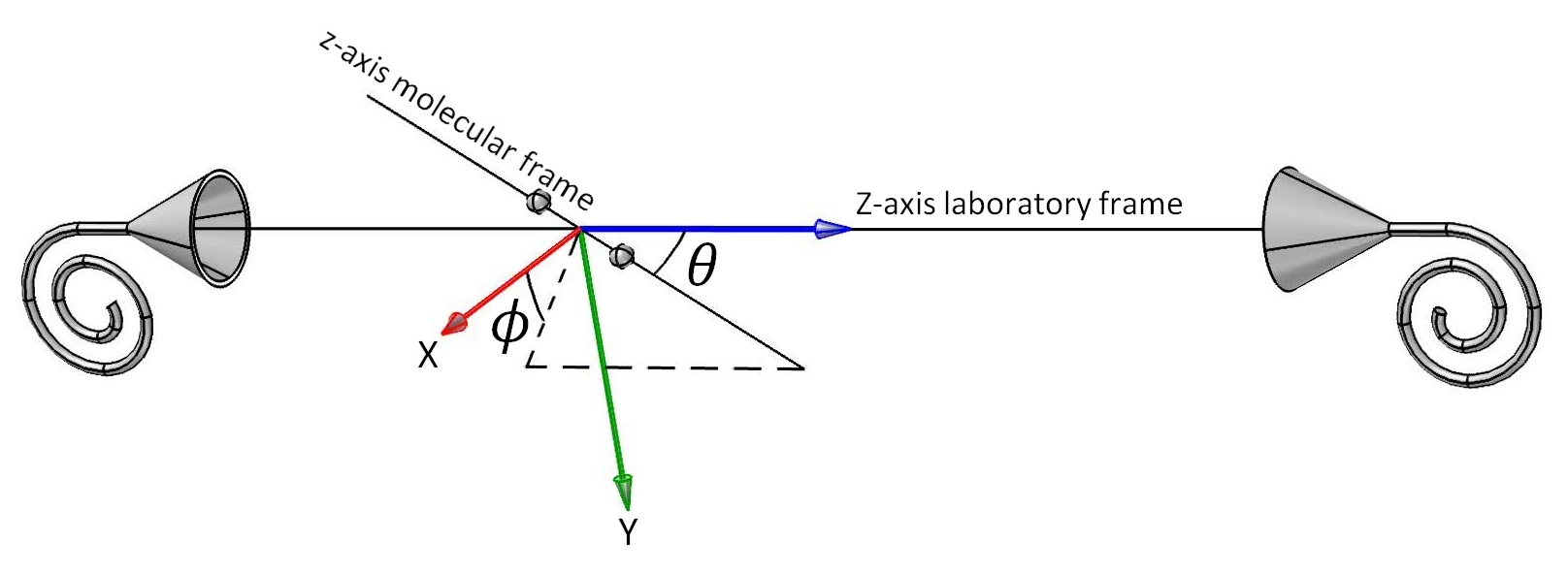}
\caption {Z-axis in laboratory frame is defined by the line crossing the center of both detectors. The variables ($\theta,\phi$) are the angles of the nuclear axis in respect to the laboratory frame and, as it is mentioned in the text, ($\Theta,\Phi$) corresponds to the detector's solid angle $\Omega$.}
\label{detector_axes}
\end{figure}

The geometry of the detection system is depicted in Fig.\ref{detector_axes}. The detectors should be aligned to ensure that if one atom is detected, the other one, arriving from the same dissociation, is also detected by the other detector.

The propagation of the fragments occurs along the internuclear axis, so that the aperture of the detectors filters a finite solid angle $\Omega$ of the possible angular positions $(\theta,\phi).$ This finite aperture effect is of crucial importance, as it provides different weights for the angular momentum states  $N=0,2$.

 Besides, there is a unique correspondence between these momenta and the time-of-flight of the fragments to the detectors. Thus, a couple of momentum eigenstates $(k_A,k_B)$ yields well-defined arrival times $(\tau_A,\tau_B).$ As we are dealing with identical fragments, each click registered in a detector may correspond to either one of the fragments. 
 
Summing up these considerations, one may think the detectors in terms of projection operators in the momentum and angular (detector's solid angle $\Omega$) spaces,  and a transition operator from 2s to 1s electronic state. The later corresponding to the process that takes place in the detection. Therefore
\begin{eqnarray}
\hat{D}_r & = & \Biggl[ \mathbb{I}_A\otimes \biggl( \vert k_r\rangle_B \vert 1s\rangle_{BB} \langle k_r\vert {}_B\langle 2s\vert \biggr) + \nonumber \\ 
 && + \biggl( \vert k_r\rangle_A \vert 1s\rangle_{AA} \langle k_r\vert {}_A\langle 2s\vert \biggr) \otimes \mathbb{I}_B \Biggr] \times  \nonumber \\ 
&& \times \int_{\Omega}d\Omega\vert\Theta ,\Phi\rangle\langle\Theta ,\Phi\vert,
\label{op-detetor}
\end{eqnarray}
\noindent where subscript $r$ stands for right side detection. For example, when $\hat{D}_r$ acts on atom B, its linear moment $k$ is detected. For the left side one has a similar expression for $\hat{D}_l$. Detection in coincidence corresponds to the application of both operators $\hat{D}_{l}\hat{D}_{r}$.  

The counting number in each detector (the probability per unit of momentum) is proportional to $P_i= \langle\Psi_{asy}\vert\hat{D}_i^{\dag}\hat{D}_i\vert\Psi_{asy}\rangle$, where $i = l$ or $r$ depending on each side is addressed. Analogously the detection in coincidence is proportional to $$P_{coinc}= \langle\Psi_{asy}\vert\hat{D}_{l}^{\dag}\hat{D}_{r}^{\dag}\hat{D}_{l}\hat{D}_{r}\vert\Psi_{asy}\rangle = |\hat{D}_l\hat{D}_r\Psi_{asy}^P|^2.$$
As we consider the para-hydrogen with $N=0$ and $N=2$, we have from Eqs.~(\ref{Psi-asymp}) and (\ref{op-detetor}):
\begin{align}
\hspace{-2cm}
\label{ProbDeDdNaCoincidenciaCos}
&|\hat{D}_l\hat{D}_r\Psi_{asy}^P|^2= \nonumber \\
& 2 \Bigg(  f_0^2(k_r)\int_{\Omega_{\cap}} d\Omega \langle{0,0}|\Theta, \Phi \rangle \langle\Theta, \Phi |0,0 \rangle + \nonumber \\ 
&+ f_2^2(k_r)\sum_{M_2,M_2^{\prime}} \int_{\Omega_{\cap}} d\Omega \langle{2,M_2^{\prime}}| \Theta,\Phi \rangle \langle{\Theta, \Phi}| 2,M_2 \rangle + \nonumber \\ 
& -2 \Big(f_2(k_r)f_0(k_r){\rm cos}(2(\delta_2-\delta_0)) \times \nonumber \\ 
& \times \sum_{M_2} \int_{\Omega_{\cap}} d\Omega \langle 0,0 | \Theta,\Phi \rangle \langle\Theta, \Phi | 2,M_2 \rangle \Big) \Bigg), \nonumber \\
\end{align}
where $\Omega_{\cap}$ is the solid angle of the detector that is farthest from the dissociation region. For this reason, the integral of the expression
Eq.(\ref{ProbDeDdNaCoincidenciaCos}) must be performed for the smaller solid angle between the two detectors. 

In the case of simple detection, one can show that
\begin{equation}
|\hat{D}_r\Psi_{asy}|^2 = \frac{\Omega}{2\pi}\sum_{N,M_N} f_N^2(k_r),
\label{result}
\end{equation}
which reflects the fact that one has no privileged axis of detection. Unlike that, in coincidence measurement one has a privileged axis formed by the line of the two detectors. This leads that the integral of the third term in Eq.(\ref{ProbDeDdNaCoincidenciaCos}) does not vanish. 
The first two terms of Eq.(\ref{ProbDeDdNaCoincidenciaCos}) are similar to the ones obtained in the simple detection case. On the other hand, the third one corresponds to an {\it interference term} between the two possible paths in the dissociation of the molecule.

The phase-shifts in Eq.~(\ref{ProbDeDdNaCoincidenciaCos}) were obtained by modifying a code written by L. F. Canto \cite{Canto} aimed at nuclear systems and changed to systems of molecular physics. In this code $\delta_0$ and $\delta_2$ are obtained as a function of the kinetic energy of the fragments, E(eV).

For the numerical analysis of the coincidence detection, we consider that the two detectors are aligned with respect to the dissociation region and that the farthest detector is at 21 cm from this region, while the nearest detector is 18 cm away from the dissociation region. As the detectors are aligned, the solid angle $\Omega_{\cap}$ considered will be the one associated with the detector that is farther from the dissociation region (at 21 cm).

The connection between the relative momentum $k$ and the difference in time-of-flights $\Delta t$ of each atom is given by $k=2(\mu/\hbar)(\Delta l/\Delta t),$
where $\Delta l=l_l-l_r$ is the modulus of the difference between the distances travelled by the atoms from the dissociation region until their detections by the detectors $\hat{D}_l$ and $\hat{D}_r$, respectively. 

In order to conciliate the theory, expressed by $|\hat{D}_e\hat{D}_d\Psi_{asy}|^2$, with the experimental counting rate, the following transformation is necessary:
\begin{equation}
|\hat{D}_e\hat{D}_d\Psi_{asy}|^2dk=h_c(\Delta t)d\Delta t.
\label{h_c(tof)}
\end{equation}

\begin{figure}[h]
\includegraphics[scale=.3]{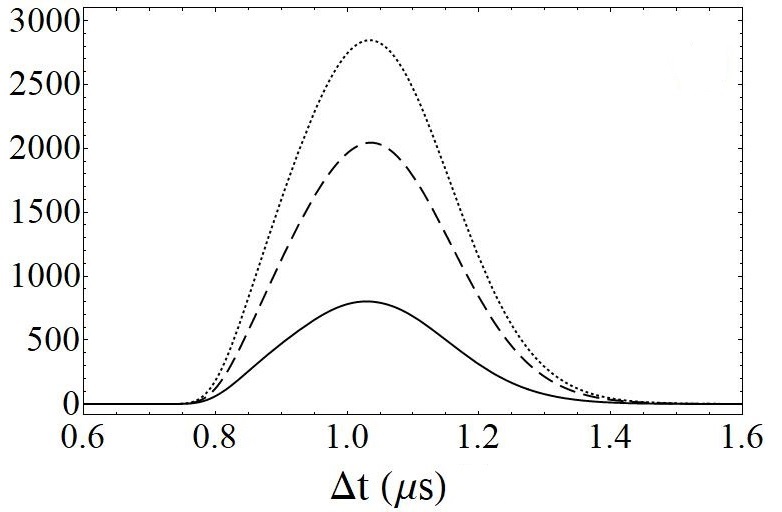}
\caption{Detection in coincidence, in arbitrary units, as a function of $\Delta t$. It is shown the comparison between the probability density of detection $h_c(\Delta t)$ (solid line, Eq.~(\ref{h_c(tof)})), the probability density of detection without interference term (dotted line), and the module of the interference term (dashed  line). Both detectors are aligned relative to the dissociation region, with one of them spaced 18 cm apart from this region and the other one 21 cm apart.}
\label{det-coinc-tof}
\end{figure}

In Fig.~\ref{det-coinc-tof} it is shown $h_c(\Delta t)$, in  arbitrary units, obtained from Eq.(\ref{h_c(tof)}), together with the contribution originated from the two first terms of Eq.~(\ref{ProbDeDdNaCoincidenciaCos}) (dotted line) and the modulus of the third one, the interference term (dashed line). The main difference between the probability density with and without interference is related to the amplitude, although the two distributions are slightly shifted.

Despite the presence of the interference term, it is not possible to observe oscillatory behaviour in $h_c(t)$, as shown in Fig.~\ref{det-coinc-tof}. This is due to the fact that the oscillations which arise from cos(2($\delta_2-\delta_0))$ are quite small in the energy range of interest, fixed by the Franck-Condon region.

\begin{figure}[h]
\begin{minipage}{8cm}
\vspace{.5cm}
\includegraphics[scale=.29]{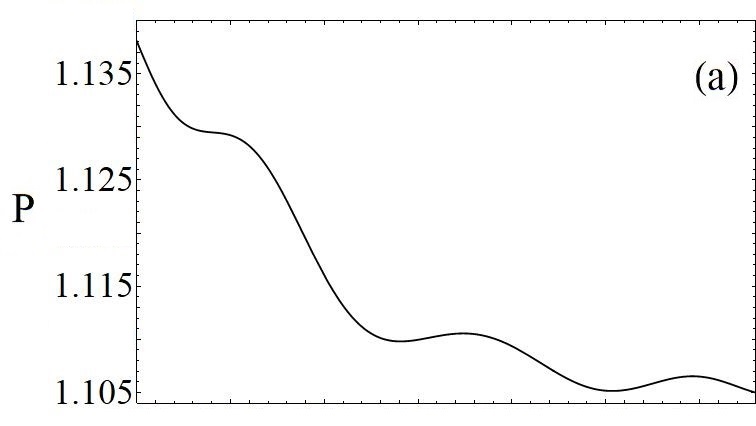}
\end{minipage}
\begin{minipage}{8cm}
\includegraphics[scale=.3]{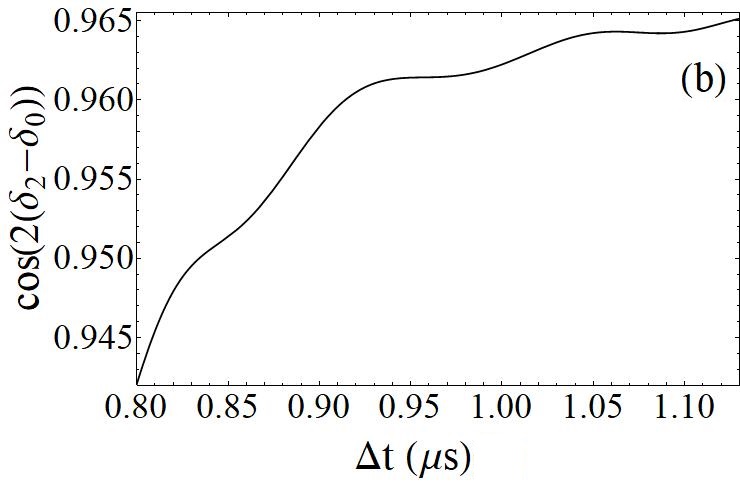}
\end{minipage}
\caption{(a) Oscillation of the normalized counting rate P($\Delta t$); (b) cos$(2(\delta_2-\delta_0))$ as a function of $\Delta t$. For details, see the text.}
\label{DetecVisibilidadeCosEmt}
\end{figure}

To enhance the phase-shift difference effect in the interference term of Eq. (\ref{ProbDeDdNaCoincidenciaCos}), we divide the expression of Eq. (\ref{ProbDeDdNaCoincidenciaCos}) by
\begin{align}
& p(k_r)=2 \Bigg(  f_0^2(k_r)\int_{\Omega_{\cap}} d\Omega \langle{0,0}|\Theta, \Phi \rangle \langle\Theta, \Phi |0,0 \rangle + \nonumber \\ 
& + f_2^2(k_r)\sum_{M_2,M_2^{\prime}} \int_{\Omega_{\cap}} d\Omega \langle{2,M_2^{\prime}}| \Theta,\Phi \rangle \langle{\Theta, \Phi}| 2,M_2 \rangle + \nonumber \\ 
& -2 \Big(f_2(k_r)f_0(k_r)\sum_{M_2}\int_{\Omega_{\cap}} d\Omega \langle 0,0 | \Theta,\Phi \rangle \langle\Theta, \Phi | 2,M_2 \rangle \Big) \Bigg),
\label{p(k_r)}
\end{align}
and then convert the result in terms of $\Delta t$. The behaviour of the resulting expression, P($\Delta t$), is displayed in Fig.~\ref{DetecVisibilidadeCosEmt}(a). It is worth mentioning that Eq.~(\ref{p(k_r)}) is equal to Eq.~(\ref{ProbDeDdNaCoincidenciaCos}), except for the replacement of cos(2($\delta_2-\delta_0))$ by 1. The quantity ${\rm P}(\Delta t)={\rm p}(\Delta t)/{\rm h}_c(\Delta t)$ plays the role of a normalized counting rate and allows us to see oscillations from the result of the detection in coincidence. 

It is important to notice that this {\it normalization} does not introduce oscillatory behaviour in P($\Delta t$), depicted in Fig.~\ref{DetecVisibilidadeCosEmt}(a), since $p(k_r)$ does not contain the information carried by the phase-shifts. In a way, by doing the division of $|\hat{D}_l\hat{D}_r\Psi_{asy}^P|^2$  by $p(k_r)$, we  remove from the data all information concerning every aspect of the system other than the phase-shift difference effect. In fact, comparing P($\Delta t$) with cos(2($\delta_2-\delta_0))$ in Fig.\ref{DetecVisibilidadeCosEmt}, we can see that their oscillations are completely connected.

In this paper we have shown that, through the coincidence detection of the H$(2s)$ atoms coming from the same H$_2$ molecule dissociation, it is possible to observe interference pattern due to different possible fragmentation paths in the dissociation process; this behaviour is connected with the phase-shifts imprinted by the repulsive effective molecular potential and reveals the coupling between the angular momentum and the vibrational states of the molecule. This result is foreseen for the data obtained in an ordinary coincidence time-of-flight detection experiment. It is also verified that this behaviour cannot be observed in an experiment involving only a single detector.

We thank L. F. Canto for his assistance with the application of his code to compute the phase-shifts. One of the authors (GJ) is thankful to J. Robert for the useful discussions on the theoretical description of detection. This work was partially supported by the Brazilian agencies CNPq,  CAPES, and FAPERJ.

\bibliographystyle{apsrev4-1}

\end{document}